# Transition Matrix Elements for Pion Photoproduction

Mohamed E. Kelabi[1]

**Abstract**
We have obtained the transition matrix elements for pion photoproduction by considering the number of gamma matrices involved. The approach based on the most general conditions of gauge invariance, current conservation and transversality. The approach is fairly consistent with literatures.

## 1.1 Introduction

The processes of scattering can change both amplitudes and phase shifts of the outgoing waves. More precisely, the effect of the scatterer is to cause transition of a particle from an initial state to another, final, state. In the case of pion photoproduction via nucleon,

$$\gamma(K) + N(P_1) \rightarrow N(P_2) + \pi(Q), \qquad (1.1)$$

the scattering processes including absorption and decay can be described, in terms of scattering matrix [1], following the convention of Brojken and Drell [2], giving

$$S = -\frac{i}{(2\pi)^2} \delta^4(P_2 + Q - K - P_1) \left( \frac{m_N^2}{4 E_2 \omega_q k E_1} \right)^{1/2} T \qquad (1.2)$$

where
$\quad P_2$ = Final nucleon four-vector
$\quad Q$ = Pion four-vector
$\quad K$ = Photon four-vector
$\quad P_1$ = Initial nucleon four-vector
$\quad m_N$ = Nucleon rest mass
$\quad E_1$ = Initial nucleon energy
$\quad k$ = Magnitude of the initial momentum
$\quad \omega_q$ = Pion energy
$\quad E_2$ = Final nucleon energy
$\quad T$ = The transition matrix element.

---

1  Physics Department, Al-Fateh University, Tripoli, Libya

The transition matrix element *T*, is related to the electromagnetic currents $J_\mu$ by the relation [3]

$$T = \varepsilon^\mu {}_{out}\langle P_2 Q | J_\mu | P_1 \rangle_{in} \qquad (1.3)$$

where $\varepsilon^\mu$ is the photon polarization vector. Eq.(1.3) can be written in general as

$$T = \bar{u}(P_2)\varepsilon^\mu \mathcal{M}_\mu u(P_1) \qquad (1.4)$$

here $u(P_1)$ is the spinor of the initial nucleon, $\bar{u}(P_2)$ is the spinor adjoint of the final nucleon, and $\mathcal{M}_\mu$ is a four-vector spinor operator, whose form remains to be determined.

## 1.2   Approach and formalism

It is convenient to rewrite the transition amplitudes, Eq. (1.4), in the most general form, allowed by Lorentz invariance

$$\varepsilon^\mu \mathcal{M}_\mu = \sum_i N_i(P,K,Q,\varepsilon) A_i(s,t,u) \qquad (1.5)$$

where $A_i(s,t,u)$ are Lorentz invariant scalar amplitudes; *s*, *t*, and *u* being the Mandelstam variables [4], and $N_i(P,K,Q,\varepsilon)$ are independent spinors which must satisfy several general conditions:
a) Must be linear in $\varepsilon$, which is the polarization vector associated with the external photon.
b) They can depend only on three four-momenta, chosen [5], [6] to be

$$P = \frac{1}{2}(P_1 + P_2) \qquad K \qquad Q$$

since the fourth can be eliminated by the conservation of energy-momentum.
c) Must be pseudoscalar under parity, to take account of the negative intrinsic parity of the pion. In effect, a factor of $\gamma_5$ must be added to each one involved.
d) They must satisfy the constraints of gauge invariance.

If we ignore for the moment the gauge invariance, then it is convenient to classify the independent spinors $N_i$ by the number of $\gamma$-matrices involved:
1) Two $\gamma$-matrices corresponds to one possibility

$$\gamma_5 \gamma\cdot\varepsilon\, \gamma\cdot K$$

since the other possibilities $\gamma_5 \gamma \cdot K \gamma \cdot K P \cdot \varepsilon = \gamma_5 P \cdot \varepsilon K^2$, where the scalar $K^2$ can be absorbed into the invariant amplitude $A_i$.

2) One $\gamma$-matrix corresponds to four possibilities

$$\gamma_5 \gamma \cdot \varepsilon \qquad \gamma_5 \gamma \cdot K K \cdot \varepsilon \qquad \gamma_5 \gamma \cdot K P \cdot \varepsilon \qquad \gamma_5 \gamma \cdot K Q \cdot \varepsilon$$

since by the conservation of energy and momenta $\gamma \cdot K + \gamma \cdot P_1 = \gamma \cdot P_2 + \gamma \cdot Q$ the other possibilities involving $\gamma \cdot P_1$ and $\gamma \cdot P_2$ can be eliminated by using Dirac equations [7]

$$\slashed{P_1} u(p_1) = m_N u(p_1) \qquad \bar{u}(p_2)\slashed{P_2} = m_N \bar{u}(p_2)$$

where we have introduced the Feynman notation [8], $\slashed{P} = \gamma_\mu P^\mu$.

3) No $\gamma$-matrices, this corresponds to three possibilities

$$\gamma_5 P \cdot \varepsilon \qquad \gamma_5 K \cdot \varepsilon \qquad \gamma_5 Q \cdot \varepsilon .$$

Any other possibilities involving more than two $\gamma$-matrices are impossible since there can only be one $\gamma$-matrix $\gamma \cdot \varepsilon$ and two $\gamma$-matrices $\gamma \cdot K \gamma \cdot K = K^2$. So the most general form allowed in Eq. (1.5), ignoring gauge invariance, can be chosen to be

$$\varepsilon^\mu \mathcal{M}_\mu = \sum_i N_i(P,K,Q,\varepsilon) B_i(s,t,u) . \qquad (1.6)$$

Here $B_i$ are scalar functions which coincide with those introduced by Ball [9]. Thus, a suitable set of $N_i$ can be written in terms of eight pseudovectors [3]

$$\begin{aligned}
N_1 &= i\gamma_5 \slashed{\varepsilon}\slashed{K} & N_5 &= -i\gamma_5 \slashed{\varepsilon} \\
N_2 &= 2i\gamma_5 P \cdot \varepsilon & N_6 &= i\gamma_5 \slashed{K} P \cdot \varepsilon \\
N_3 &= 2i\gamma_5 Q \cdot \varepsilon & N_7 &= i\gamma_5 \slashed{K} K \cdot \varepsilon \qquad (1.7)\\
N_4 &= 2i\gamma_5 K \cdot \varepsilon & N_8 &= i\gamma_5 \slashed{K} Q \cdot \varepsilon
\end{aligned}$$

where the factors $i$ and 2 are introduced for convenience [5]. Apply the conditions due to gauge invariance, current conservation $K^\mu \mathcal{M}_\mu = 0$, on Eq. (1.6) then transversality $K \cdot \varepsilon = 0$ gives $N_4 = N_7 = 0$, hence one can set $B_4 = B_7 = 0$. On the other hand, because of transversality, $\gamma \cdot K$ anticommutes with $\gamma \cdot \varepsilon$, this leads to

$$2i\gamma_5(P\cdot KB_2 + Q\cdot KB_3) - i\gamma_5 \slashed{K}(B_5 - P\cdot KB_6 - Q\cdot KB_8) = 0$$

where we have imposed the condition $K^2 = 0$. Following the literatures, defining [10]:

$$A_1 = B_1 - m_N B_6$$
$$A_2 = \frac{B_3}{P\cdot K} = -\frac{B_2}{Q\cdot K}$$
$$A_3 = -B_8$$
$$A_4 = -\frac{B_6}{2}$$

which can be transposed on the form

$$\begin{aligned} B_1 &= A_1 - 2m_N A_4 \\ B_2 &= -Q\cdot K\, A_2 \\ B_3 &= P\cdot K\, A_2 \\ B_5 &= -2P\cdot K\, A_4 - Q\cdot K\, A_3 \\ B_6 &= -2A_4 \\ B_8 &= -A_3 \end{aligned} \quad (1.8)$$

to finally obtain

$$\varepsilon^\mu \mathcal{M}_\mu = \sum_{i=1}^{4} M_i(P,K,Q,\varepsilon) A_i(s,t,u). \quad (1.9)$$

Substituting Eq. (1.8) back into Eq. (1.6), and making use of Eq. (1.7), we have

$$\begin{aligned} \varepsilon^\mu \mathcal{M}_\mu =\ & i\gamma_5 \slashed{\varepsilon}\slashed{K}(A_1 - 2m_N A_4) + 2i\gamma_5 P\cdot\varepsilon(-Q\cdot K\, A_2) \\ & + 2i\gamma_5 Q\cdot\varepsilon(P\cdot K\, A_2) - i\gamma_5\slashed{\varepsilon}(-2P\cdot K\, A_4 - Q\cdot K\, A_3) \\ & + i\gamma_5\slashed{K}P\cdot\varepsilon(-2A_4) + i\gamma_5\slashed{K}Q\cdot\varepsilon(-A_3). \end{aligned} \quad (1.10)$$

Rearranging Eq. (1.10) by factoring out $A_i$, and comparing with Eq. (1.9) of the form,

$$\varepsilon^\mu \mathcal{M}_\mu = M_1 A_1 + M_2 A_2 + M_3 A_3 + M_4 A_4, \quad (1.11)$$

obtaining the coefficients of $A_i$,

$$M_1 = i\gamma_5 \not{\varepsilon} \not{K}$$
$$M_2 = 2i\gamma_5 (P \cdot K Q \cdot \varepsilon - P \cdot \varepsilon Q \cdot K)$$
$$M_3 = i\gamma_5 (\not{\varepsilon} Q \cdot K - \not{K} Q \cdot \varepsilon) \quad (1.12)$$
$$M_4 = 2i\gamma_5 (\not{\varepsilon} P \cdot K - \not{K} P \cdot \varepsilon - m_N \not{\varepsilon} \not{K}),$$

the required relativistic invariant operators $M_i$ [5], [9].

It is remaining to mention that, the invariant amplitudes $A_i$ can be extracted [11] from the Born terms [12], [13], shown in Figure 1-1, then comparing the results with Eq. (1.11).

$\varepsilon^\mu \mathcal{M}_\mu$ = 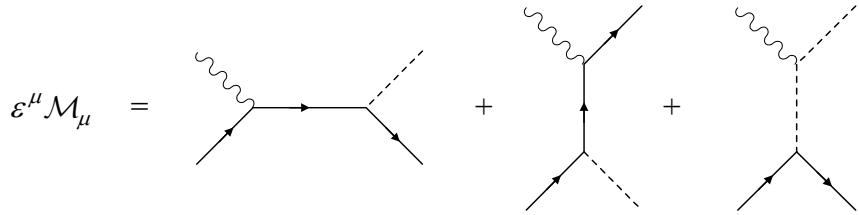

Figure 1-1. Born terms for pion photoproduction.

## 1.3 Conclusion

The techniques used here based on the most general conditions of gauge invariance, current conservation, and transversality condition. The calculated relativistic invariant operators make it clear to work with transition matrix for pion photoproduction, and hence to extract the required invariant amplitudes in a simple straightforward way.